\documentclass[prd, aps, showpacs]{revtex4}

\usepackage{latexsym, graphicx}

\begin{document}

\title{New Insights into 
Uniformly Accelerated Detector in a Quantum Field}
\author{Shih-Yuin Lin
\footnote{Current address: Physics Division, National Center for
Theoretical Sciences, National Tsing-Hua University, Hsinchu 30013,
Taiwan}} \email{sylin@phys.cts.nthu.edu.tw} \affiliation{Center for
Quantum and Gravitational Physics, Institute of Physics,\\ Academia
Sinica, Nankang, Taipei 11529, Taiwan}
\author{B. L. Hu}
\email{blhu@umd.edu} \affiliation{Department of Physics, University
of Maryland, College Park, Maryland 20742-4111, USA}
\date{September 30, 2006}

\begin{abstract}
We obtained an exact solution for a uniformly accelerated Unruh-DeWitt 
detector interacting with a massless scalar field in (3+1) dimensions 
which enables us to study the entire evolution of the total system, 
from the initial transient to late-time steady state. 
We find that the Unruh effect as derived from time-dependent
perturbation theory is valid only in the transient stage and is
totally invalid for cases with proper acceleration smaller than the
damping constant. We also found that, unlike in (1+1)D results, the
(3+1)D uniformly accelerated Unruh-DeWitt detector in a steady state 
does emit a positive radiated power of quantum nature at late-times, 
but it is not connected to the thermal radiance experienced by the 
detector in the Unruh effect proper.\\
\\
{\it -- Invited talk given by SYL at the conference of International
Association for Relativistic Dynamics (IARD), June 2006, Storrs,
Connecticut, USA.}
\end{abstract}

\maketitle

\section{Introduction}

A uniformly accelerated detector (UAD) moving in Minkowski vacuum
experiences a thermal bath at the temperature $T_U=\hbar a/2\pi
ck_B$ with the proper acceleration $a$ \cite{Unr76}. This effect is
called the Unruh effect and the temperature $T_U$ is called the
Unruh temperature.

The Unruh effect was orginally derived and is usually demonstrated
under the framework of time-dependent perturbation theory
\cite{Unr76, DeW79, BD}. Consider a point-like quantum mechanical
object, the ``detector", with internal degree of freedom $Q$
coupling to a quanum field $\Phi$ through the interacting
Hamiltonian $H_I = \lambda_0 Q(\tau) \Phi(z^\mu(\tau))$, where
$\lambda_0$ is the coupling constant, $\tau$ is the proper time
ofthe detector and $z^\mu(\tau)$ is the trajectory the uniformly
accelerated detector is going along. Suppose at the initial moment
$\tau_0$ the initial state for the detector-field system can be
factorized into
\begin{equation}
  |\left.\tau_0 \right> = |\left. E_0\right> \otimes
  |\left. 0_M\right>\label{initstat}
\end{equation}
where $|\left. E_0\right>$ is the ground state of the free detector
and $|\left. 0_M\right>$ is the Minkowski vacuum of the free field.
Then, from time-dependent perturbation theory in quantum mechanics,
to first order in $\gamma \sim \lambda_0^2$, the transition
probability from the ground state to the n-th excited state of the
detector is given by \cite{BD}
\begin{eqnarray}
  P_{0\to n} &=&
  {\lambda_0^2\over 2\pi\hbar^2} \int_{-\infty}^\infty d\tau {(E_n-E_0)
  \left|\left< E_n \right.| Q(0)|\left. E_0\right>\right|^2 \over
   e^{2\pi (E_n-E_0)/a\hbar} -1}. \label{Planck}
\end{eqnarray}
which is non-zero. In particular, for a simple harmonic oscillator
detector with natural frequency $\Omega_r$, all the transition
probabilities with $n>1$ are $O(\gamma^2)$, and the only
non-vanishing $P$ of $O(\gamma)$ is
\begin{equation}
  P_{0\to 1} = {\lambda_0^2\over 4\pi m_0} {\eta \over
   e^{2\pi \Omega_r /a} -1}
\label{Psho}
\end{equation}
where $\eta \equiv \int_{-\infty}^\infty d\tau$ is the duration of
interaction in the detector's proper time. Accordingly one claims
that a uniformly accelerated detector moving in Minkowski vacuum
experiences the same effect as does an inertial detector immersed in
a thermal bath at the Unruh temperature % $T_U = \hbar a/2\pi k_B$,
(which can be read off from the Planck factors in $(\ref{Planck})$
and $(\ref{Psho})$). When $a=0$, the transition probability per unit
time $P_{0\to 1}/\eta$ vanishes, which implies that there is no
excitation in an inertial detector initially prepared in its ground
state \cite{DeW79}.

Recently we studied an Unruh-DeWitt detector theory in (3+1)D and
obtained a complete description of the combined system with exact
expressions for the evolution of the detector and field
correlations\cite{LH2005, LH06a}. With these non-perturbative
results, we found that some long-held beliefs based on the
perterbation theory are not true. Furthermore, some intuitions
gained from (1+1)D results cannot be applied to (3+1)D case, though
the spacetime dimension does not matter in the above arguments
focused on the response of the detector. We summarize these points
in the following.

%%%%%%%%%%%%%%%%%%%%%%%%%%%%%%%%%%%%%%%%%%%%%%%%%%%%%%%%%%%%%%%%%%%%%%
\section{The Model}

We consider the combined system of a Unruh-DeWitt (UD) detector
interacting with a massless scalar field in (3+1)D Minkowski space,
described by the action \cite{Unr76,DeW79, LH2005}
\begin{equation}
  S = \int d\tau {m_0\over 2}\left[ \dot{Q}^2 -\Omega_0^2 Q^2\right]
    -\int d^4 x {1\over 2}\partial_\mu\Phi \partial^\mu\Phi
    +{\lambda_0} \int d\tau \int d^4 x Q(\tau) \Phi (x)
  \delta^4\left(x^{\mu}-z^{\mu}(\tau)\right) . \label{Stot1}
\end{equation}
Here $Q$ is the internal degree of freedom of the detector, assumed
to be a harmonic oscillator with mass $m_0$ and natural frequency
$\Omega_0$, $\tau$ is the detector's proper time, and $\dot{Q}
\equiv dQ(\tau)/d\tau$. $\Phi$ is the massless scalar field, and
$\lambda_0$ is the coupling constant.

For simplicity, we do not consider the trajectory of the detector
$z^{\mu}$  as a dynamical variable (for a discussion of the case
where the trajectory is determined by its interplay with the quantum
field, see, \cite{JH1,JHIARD}), but gauged by an external agent. We
assume the UD detector is moving in a prescribed trajectory in
uniform acceleration: $z^\mu(\tau)=( a^{-1}\sinh a\tau, a^{-1}\cosh
a\tau,0,0)$ with $x^0-x^1=0$ being the event horizon for the
detector.
%We are thus dealing with a hybrid of quantum field theory
%of the massless scalar field, quantum mechanics of the detector's
%intenal degree of freedom, and the classical external agent.

When $a=0$, the UD detector theory is a special case of the harmonic
oscillator quantum Brownian motion (QBM) model, studied before
by many authors (see references in e.g., \cite{HPZ}).  The relation
between these two models becomes clear when we make the substitutions 
$Q$ for $x$, $\int d^3 k$ for $\sum_n$, $\Phi_k$ for $q_n$ and $-
\lambda_0 e^{ikz}$ for $C_n$ in \cite{HPZ}. The QBM model incorporates 
the effect of the environment (the quantum field) on the system 
(harmonic oscillator) with dissipative and stochastic dynamics. 
This shows that even for the $a=0$ case the detector is not just laying
idle but has interesting physical behaviors due to its interaction
with the fluctuations in the quantum field.

\section{Exact Evolution of Operators}

%To compare with the conventional argument in Section I,
We start at the initial time $\tau=\tau_0$ with the same initial
state $(\ref{initstat})$ and assume the initial operators are those
for free theories. Suppose the coupling $\lambda_0$ is not turned on
until $\tau_0$, when all the dynamical variables are allowed to
interact and evolve. By virtue of the linear coupling between
$\hat{Q}$ and $\hat{\Phi}$ in $(\ref{Stot1})$, the time evolution of
$\hat{\Phi}({\bf x})$ and $\hat{Q}$ from the Heisenberg equations of
motion is simply a linear transformation in the phase space spanned
by $( \hat{\Phi}({\bf x}),\hat{\Pi}({\bf x}) , \hat{Q},\hat{P})$.
Thus $\hat{\Phi}(x)$ and $\hat{Q}(\tau)$ can be expressed in the
form
\begin{eqnarray}
  \hat{\Phi}(x) &\sim& \int d^3{\bf k}
    \left[f^{(+)}(x;{\bf k})\hat{b}_{\bf k} +
    f^{(-)}(x;{\bf k})\hat{b}_{\bf k}^\dagger
    \right] + %\sqrt{\hbar \over 2\Omega_r m_0}\left[
    f^a(x)\hat{a}+ f^{a*}(x)\hat{a}^\dagger ,\label{phiab}\\
  \hat{Q}(\tau) &\sim& \int d^3{\bf k}
  \left[q^{(+)}(\tau,{\bf k})\hat{b}_{\bf k} +
    q^{(-)}(\tau,{\bf k})\hat{b}_{\bf k}^\dagger\right] +
    %\sqrt{\hbar\over 2\Omega_r m_0}\left[
    q^a(\tau)\hat{a}+q^{a*}(\tau)\hat{a}^\dagger ,\label{qab}
\end{eqnarray}
where ($\hat{b}_{\bf k}^\dagger$, $\hat{b}_{\bf k}$) and
($\hat{a}^\dagger$, $\hat{a}$) are the creation and annihilation
operators defined in free theories for the scalar field and the
detector, respectively. The whole problem is now transformed from
solving the Heisenberg equations of motion for the operators into
one of solving for the c-number functions $f^s(x)$ and $q^s(\tau)$
with suitable initial conditions.

After the regularization and renormalization of the retarded Green's
function, similar in spirit to that in deriving the
Abraham-Lorentz-Dirac equation for moving electrons in classical
electrodynamics \cite{JH1} \footnote{Since the UD detector
considered here is a quantum mechanical object, there is a natural
cutoff on the frequency at the energy threshold for the creation of
detectors. Thus it is justified to assume that the detector has a
small but finite extent in spacetime.}  the back reaction of the
quantum field is incorporated into the equation of motion for
$q^{(+)}$, which reads
\begin{equation}
  (\partial_\tau^2 +2\gamma\partial_\tau+ \Omega_r^2)q^{(+)}
  (\tau;{\bf k}) = {\lambda_0\over m_0} f^{(+)}_0(z(\tau);{\bf k}),
  \label{eomq2}
\end{equation}
where $f_0^{(+)}(x;{\bf k})\equiv  \exp (-i\omega t+i{\bf k\cdot
x})$ is the free field solution in Minkowski coordinate, $\Omega_r$
is the renormalized natural frequency and  $\gamma \equiv
\lambda_0^2/ 8\pi m_0 $ is the damping constant resulting from the
interaction with the field. We see that  $q^{(+)}$ behaves like a
driven damped harmonic oscillator with dissipation induced by the
vacuum fluctuations of the scalar field. Eq.$(\ref{eomq2})$ is
causal and local in $\tau$. Once the form of $f^{(+)}_0$ is given,
$q^{(+)}(\tau)$ in $(\ref{eomq2})$ is totally determined by the
motion of the detector from $\tau_0$ to $\tau$. In other words, the
response of $q^{(+)}$ here is purely kinematic.

As for the $q^a$ coefficient of $\hat{a}$, its equation of motion
including the back reaction of the field looks similar. $q^a$ acts
like a damped harmonic oscillator with the renormalized natural
frequency $\Omega_r$ and the damping constant $\gamma$ but without
the driving force.

%Since the equations of motion for $q^{(+)}(\tau)$ and $q^a(\tau)$
%are derived by a coarse-graining procedure @??@, the evolution of
%them are ``exact" in the sense that the exact solutions of those
%equations are well-defined up to the precision of this theory
%due to the natural frequency cutoff.

\section{Internal activities of the detector}

For the detector-field system initially prepared in the factorized
initial state $(\ref{initstat})$, the two-point functions of $Q$
will split into two parts,
$\left<\right.Q(\tau)Q(\tau')\left.\right> = \left<\right.E_0 \,|\,
E_0\left.\right>\left< \right. Q(\tau) Q(\tau')\left.\right>_{\rm
v}+ \left< \right.Q(\tau)Q(\tau')\left. \right>_{\rm a} \left< 0_M|
0_M\right>$. Here $\left< \right. Q(\tau)Q(\tau')\left.\right>_{\rm
v}$ can be interpreted as accounting for the response to the vacuum
fluctuations of the quantum field, while $\left< \right.
Q(\tau)Q(\tau')\left.\right>_{\rm a}$ corresponds to the intrinsic
quantum fluctuations in the detector.

The two-point functions of the detector with respect to the vacuum,
$\left<\right. Q(\eta)Q(\eta')\left.\right>_{\rm v} \sim \int
d^3{\bf k}q^{(+)}(\tau;{\bf k}) q^{(-)}(\tau;{\bf k})$ with
$\eta \equiv \tau -\tau_0$ being the duration of interaction, 
can be explicitly obtained from the solution of $q^{(+)}$. The
coincidence limit of it looks like
\begin{equation}
  \left<\right. Q(\eta)^2\left.\right>_{\rm v}
  = {\hbar\lambda_0^2 \theta(\eta)\over (2\pi m_0 \Omega)^2}
   \left[ \Lambda_0 e^{-2\gamma\eta}\sin^2\Omega\eta
  + ({\rm regular\,\,terms})\right],
\end{equation}
where $\Lambda_0 \sim -\ln |\tau'_0-\tau_0|$ is finite in real
processes because $|\tau'_0-\tau_0|$ characterizes the time scale
that the interaction is turned on. (This means that $\Lambda_0$
would not be important at late times: for every finite value of
$\Lambda_0$, the $\Lambda_0$-term vanishes as
$\gamma\eta\to\infty$.) In Ref. \cite{LH2005}, the evolution of the
regular part of $\left<\right.Q(\eta)^2\left.\right>_{\rm v}$ has
been shown. Roughly speaking it saturates exponentially in the
detector's proper time to a positive number.

The coincidence limit of the two-point function $\left<\right.
\dot{Q}(\eta)\dot{Q}(\eta')\left.\right>_{\rm v}$ reads
\begin{eqnarray}
  \left<\right. \dot{Q}(\eta)^2\left.\right>_{\rm v}
    = {\hbar\lambda_0^2 \theta(\eta)\over ( 2\pi m_0\Omega)^2}
   \left[ \Lambda_1 \Omega^2 + \Lambda_0 e^{-2\gamma\eta}\left(\Omega
   \cos\Omega\eta-\gamma \sin\Omega\eta \right)^2 + \cdots\right]
\end{eqnarray}
where $\Lambda_1 \sim -\ln |\tau-\tau'|$ corresponds to the
time-resolution of this theory. The regular part of $\left<\right.
\dot{Q}(\eta)^2\left.\right>_{\rm v}$ acts quite similarly to those for
$\left<\right.Q(\tau)^2\left.\right>_{\rm v}$.

For the expectation values of the detector two-point functions with
respect to the ground state, the coincidence limits of them are
straightforward and independent of the proper acceleration $a$. The
quantity $\left<\right.Q(\eta)^2\left.\right>_{\rm a}\sim (q^a)^*
q^a$ decays exponentially due to the dissipation of its zero-point
energy to the field. As $\left<\right.Q(\eta)^2 \left.\right>_{\rm
a}$ decays, $\left<\right.Q(\eta)^2\left.\right> _{\rm v}$ grows in
such a way that $\left<\right.Q^2\left.\right> =
\left<\right.Q^2\left.\right>_{\rm a}
+\left<\right.Q^2\left.\right>_{\rm v}$ saturates at late times. For
$\left<\right.\dot{Q}(\eta)^2 \left.\right>_{\rm a}$ and $
\left<\right. \Delta \dot{Q}(\eta)^2\left.\right>$, their behavior
are similar.

One can express the reduced density matrix $\rho^R(Q,Q')$ for the
detector in terms of the above two-point functions of the detector
to study the statistical properties of the detector such as entropy
of entanglement and purity relevant to quantum information
processing and teleportation issues \cite{LH06a}. Here we use the
reduced density matrix to compare our results obtained from exact
solutions with those from conventional perturbation theory.

The initial state ($\ref{initstat}$) implies that $\rho^R(Q,Q')$ is
a Gaussian function of $Q$ and $Q'$. Transforming $\rho^R(Q,Q')$ to
the representation in the basis of energy-eigenstate for the free
harmonic oscillator $Q$,
\begin{equation}
  \rho^R(Q,Q') = \sum_{m,n\geq 0}\rho_{m,n}^R \phi_m(Q) \phi_n(Q')
\label{rhoeigen}
\end{equation}
where $\phi_n(Q)$ is the wave function for the n-th excited state,
then the transition probability from the initial ground state to the
first excited state is given by the $m=n=1$ component,
\begin{equation}
  \rho^R_{1,1} = { \hbar \left[m_0^2\left<\right. \dot{Q}^2 \left.\right>
    \left<\right. Q^2 \left.\right> -m_0^2\left<\right. \{\dot{Q},Q\}
    \left.\right>^2-{\hbar^2\over 4 }\right]
  \over \left[\left({m_0^2\over \hbar \alpha^2}\left<\right.
    \dot{Q}^2 \left.\right>
   + {\hbar\over 2}\right) \left(\left<\right. Q^2 \left.\right>
       \hbar \alpha^2+ {\hbar\over 2} \right)-
    m_0^2\left<\right. \{ \dot{Q}, Q\} \left.\right>^2\right]^{3/2}},
\end{equation}
with $\alpha = \sqrt{m_0\Omega_r/\hbar}$ and $\{A,B\} \equiv (AB+BA)/2$. 
When $\eta\equiv\tau-\tau_0 \gg a^{-1}$, expanding $\left<\right.\cdots
\left.\right>$ in terms of $\gamma\equiv\lambda_0^2/8\pi m_0$, the
approximate value to first order in $\gamma$ is
\begin{equation}
  \rho^R_{1,1}|_{\gamma\ll 1} \stackrel{\eta\gg a^{-1}}{\longrightarrow}
  {\lambda_0^2\over 4\pi m_0}\left[ {\eta\over e^{2\pi\Omega_r/a}-1}
  + {\Lambda_1 +\Lambda_0 \over 2\pi \Omega_r} \right] 
%(*************************************************************************to be corrected***),
\label{rhopert}
\end{equation}
from the results in Ref. \cite{LH2005}. We see that the first term of
($\ref{rhopert}$) gives the transition probability $(\ref{Psho})$
but emphatically it is not in a steady state situation. The
approximation used in obtaining ($\ref{rhopert}$) is valid only at
$a^{-1} \ll \eta \ll \gamma^{-1}$, when the system is still in
transient. If $a < \gamma$, no perturbative regime exists at all. So
the $a=0$ case is beyond the reach of pertubation theory, and the
conventional wisdom from perturbation theory that  no transition
occurs in an inertial detector is simply untenable. In contrast,
$\rho^R_{1,1}$ at $a=0$ behaves quite similarly to those cases with
nonzero acceleration \cite{LH2005}. This agrees with our expectation
when we observed that the UD detector theory with $a=0$ is a special
case of the model of the quantum Brownian motion \cite{HPZ}, where
there is a great deal of interplay between the oscillator and the
quantum field.

Note further that the two additional (divergent) constants
$\Lambda_0$ and $\Lambda_1$ present in ($\ref{rhopert}$) need be
kept throughout the calculation because, if $\Lambda_1$ was
subtracted naively, the uncertainty principle will be violated or
$\sqrt{\left<\right. P^2 \left.\right>\left<\right. Q^2
\left.\right>} |_{\Lambda_1=0} < \hbar/2$ at late times for small
enough $a$. With these two divergent constants, of course, the
scenario of the transition process will be quite different from the
conventional ones. Further exposition of these new results are
contained in \cite{LH06a}.

%%%%%%%%%%%%%%%%%%%%%%%%%%%%%%%%%%%%%%%%%%%%%%%%%%%%%%%%%%%%%%%%%%%%%%%%%%%
\section{Classical and Quantum Radiation}
%%%%%%%%%%%%%%%%%%%%%%%%%%%%%%%%%%%%%%%%%%%%%%%%%%%%%%%%%%%%%%%%%%%%%%%%%%%

It is common knowledge that an accelerated point-charge coupled with
electromagnetic(EM) field give rise to EM radiation \cite{Jackson,
rohr, Boulware}. Since our accelerated detector is also a point-like
object coupled with a quantum field, it is natural to ask whether
there is radiation emitted {\it from} a UAD, 
even under steady state conditions, as opposed to
the thermal radiance experienced {\it by} the detector. Some even
view the radiation emitted from a UAD as evidence of
Unruh effect \cite{ChenTaj,Chen,Leinaas}. (For a critique of this
view and explanation, see, e.g., \cite{CapHR,CapHJ}.)

Prior work in (1+1) dimensions shows that there is no emitted
radiation from a uniformly accelerated oscillator under equilibrium
conditions (steady state and uniform acceleration) \cite{Grove}.
%while there exists a ``polarization cloud" around it \cite{MPB93}.
Nevertheless, most experimental proposals on the detection of Unruh
effect are designed for the physical four dimensional spacetime, so
one needs to examine the question for (3+1) dimensions. We have
performed such an analysis which offer some new insights on this
question.

Following a similar argument in classical theory \cite{rohr}, the
radiation power emitted by the UD detector in (3+1)D is given by
\begin{equation}
  \left<{dW^{\rm rad}\over d\tau_-}\right> = -\lim_{r\to\infty}
    \int r^2 d\Omega_{\rm II} \,\,u^\mu
    \left<\right.T_{\mu\nu}\left.\right>_{\rm ren} v^\nu .
\label{Qradformula0}
\end{equation}
Here $\tau_-(x)$ is the detector proper time at the moment when the 
spacetime point $x$ (where the retarded field is measured)  lies on the 
future lightcone with origin located at $z^\mu(\tau_-)$
(see Eq.(34) in Ref. \cite{LH2005}). 
The quantum expectation value of the renormalized stress-energy 
tensor $\left<T_{\mu\nu}\right>_{\rm ren}$ is obtained by calculating
\begin{equation}
   \left< T_{\mu\nu}[\Phi(x)]\right>_{\rm ren} = \lim_{x'\to x}
   \left[ {\partial\over \partial x^\mu}{\partial\over\partial x'^\nu}
   -{1\over 2}g_{\mu\nu} g^{\rho\sigma}{\partial\over \partial x^\rho}
   {\partial\over\partial x'^\sigma}\right] G_{\rm ren}(x,x'),
\label{expTmn}
\end{equation}
where $G_{\rm ren}$ is the renormalized two-point function of the
field, defined by $ G_{\rm ren}(x,x') \equiv\left<\right.\hat{\Phi}(x)
\hat{\Phi}(x')\left.\right> -G_{\rm v}^{00}(x,x')$ with the Green's
function for free fields $G_{\rm v}^{00}$ subtracted. After some
algebra, it turns out that $r^2 u^\mu \left<\right.T_{\mu\nu}\left.
\right>_{\rm ren} v^\nu$ is regular and non-vanishing at the null
infinity of Minkowski space ($r\to \infty$) even in steady state
\cite{LH2005}, when the radiation power $(\ref{Qradformula0})$ can
be written as
\begin{equation}
  \left<{dW^{\rm rad}\over d\tau_-}\right> =
  {\lambda_0^2\over 8\pi} \int_0^\pi d\theta \sin \theta
    \left\{\left<\right. \dot{Q}^2\left.\right>_{\rm tot}
    -{\hbar\Theta_{+-}\over \pi m_0}+
   a^2 \cos^2\theta \left<\right. Q^2\left.\right>_{\rm tot}
   + a\cos\theta \left[\left<\right. \{Q,\dot{Q}\}\left.\right>_{\rm tot}
   - {\hbar\Theta_{+X}\over \pi m_0}\right] \right\}.
\label{Qradformula1}
\end{equation}
This is the quantum (massless) scalar field radiation emitted by the
UAD in (3+1)D spacetime. The quantities in this formula are defined
in Eqs. (103), (104) and Appendix A of Ref. \cite{LH2005}.

The first term in $(\ref{Qradformula1})$, $\left<\right.\dot{Q}^2
\left.\right>_{\rm tot}$, goes to zero at late times \cite{LH2005}, 
so the corresponding monopole radiation ceases after the transient. 
The interference between the quantum radiation induced by the vacuum 
fluctuations and the vacuum fluctuations themselves totally obliterate 
any information pertaining to the appearance of the Unruh 
effect in this part of the radiation. Its behavior  is analogous to
that in (1+1)D: emitted radiation from UAD is only
associated with nonequilibrium process \cite{CapHJ}.

The total screening of the monopole radiation corresponding to
$\left<\right.\dot{Q}^2\left.\right>_{\rm tot}$ is actually a
consequence of energy conservation. We found that the energy of the
``dressed" detector

\begin{equation} E(\eta)\equiv (m_0/
2)[\left<\right. \dot{Q}^2(\eta)\left.\right>+
\Omega_r^2\left<\right.Q^2(\eta)\left. \right> ]
\end{equation}
changes in time as
\begin{eqnarray}
  -\dot{E}(\eta) &=& {\lambda_0^2\over 4\pi}
    \left<\right.\dot{Q}^2(\eta)\left.\right>_{\rm tot} ,
\label{Econserv}
\end{eqnarray}
for all $\eta >0$. This relation says that the rate of energy-loss
of the dressed detector is equal to the radiated power via the
monopole radiation corresponding to
$\left<\right.\dot{Q}^2(\eta)\left.\right>_{\rm tot}$. Thus
Eq.$(\ref{Econserv})$ is simply a statement of energy conservation
between the detector and the field. The external agent which drives
the detector along the trajectory $z^\mu(\tau)$ has no additional
influence in this channel.

At late times, while $\left<\right.\dot{Q}^2\left.\right>_{\rm tot}$
ceases, a positive radiated power flow to the null infinity of
Minkowski space still remains:
\begin{equation}
  \left<{dW^{\rm rad}\over d\tau_-}\right> \to
  {\hbar\lambda_0^2 \over 8\pi^2 m_0 }\left\{
  {a^3\over 3\Omega_r^2} -a %\right.\nonumber\\ &&\left.
  -{2\over 3}\left[ {a^3\over \Omega_r^2}-a  + 2\gamma
    + {\rm Re}\left[{i(\gamma+i\Omega)\over a\Omega}
      \left[ (\gamma+i\Omega)^2-a^2\right] \psi^{(1)}
    \left({\gamma+i\Omega \over a}\right)\right]\right] \right\}.
\label{QMrad}
\end{equation}
Thus we conclude that there exists a steady, positive radiated power
of quantum nature emitted by the detector even when the detector is
in a steady state. For large $a$, the first term in $(\ref{QMrad})$
dominates, and the radiated power is approximately
\begin{equation}
  \left<{dW^{\rm rad}\over d\tau_-}\right> \approx
  {\lambda_0^2 \over 4\pi } {a^2\over 3}{\hbar a\over 2\pi m_0\Omega_r^2}
  \propto a^2 T_U,
\end{equation}
where $T_U$ is the Unruh temperature. This could be interpreted as
a hint of the Unruh effect.

The steady radiated power flow $(\ref{QMrad})$ does not originate
from the thermal radiance that the detector experiences as in the
Unruh effect, since the internal energy of the dressed detector is
conserved only in relation to the radiated energy of a monopole
radiation corresponding to $\left<\right.\dot{Q}^2\left.\right>_{\rm
tot}$, which contributes nothing to $(\ref{QMrad})$. Learning from
the EM radiation emitted by a uniformly accelerated charge
\cite{rohr, Boulware}, we expect that the above non-vanishing
radiated energy of quantum origin is supplied by the external agent
driving the motion of the detector.

%%%%%%%%%%%%%%%%%%%%%%%%%%%%%%%%%%%%%%%%%%%%%%%%%%%%%%%%%%%%%%%%%%%
\section{Summary}
%%%%%%%%%%%%%%%%%%%%%%%%%%%%%%%%%%%%%%%%%%%%%%%%%%%%%%%%%%%%%%%%%%%

Our exact solution indicates that %Unruh effect as derived in
the conventional time-dependent perturbation theory in demonstrating 
Unruh effect is valid only in transient, with the duration of 
interaction timed between $1/a$ and $1/\gamma$. 
For the cases with proper acceleration $a$ smaller than the damping
constant $\gamma$, time-dependent perturbation theory is invalid.
Moreover, even with $a=0$ there still exists non-trivial behavior in
the detector when coupled with the quantum field. We also found new
divergent constants present in the transition probability from the
initial ground state of the detector to the excited states. They
alter the scenario about the evolution of the system.

Going outside of the detector we found that unlike the (1+1)D case,
the (3+1)D uniformly accelerated UD detector in a steady state does
emit a positive radiated power of quantum nature. When the proper
acceleration $a$ is large, this flux is approximately proportional
to the Unruh temperature $T_U$, so it could be interpreted as a hint
of the Unruh effect. Nevertheless, since the total energy of the
dressed detector is conserved only with the radiated energy of a
monopole radiation which ceases in steady state, the {\it hint}
of the Unruh effect in the late-time radiated power in (3+1)D is not
connected to the thermal radiance experienced by the detector in the
Unruh effect {\it proper}. The experiments proposed so far
\cite{Chen} for the detection of Unruh radiation were not meant to
be nor are they sensitive enough for this quantum radiation.
\\

\noindent{\bf Acknowledgments} This work is supported in part by the
NSC Taiwan under grant NSC93-2112-M-001-014 and by the US NSF under
grant PHY-0601550 and PHY-0426696.

\end{document}